\documentclass[namedreferences]{SolarPhysics}
\usepackage[optionalrh]{spr-sola-addons} 
\usepackage{graphicx}        
\usepackage{color}           
\usepackage{url}             

\begin{document}

\begin{article}

\begin{opening}
\title{Fast magnetosonic modes of  cylindrical   magnetotail}

\author{I. S.~\surname{Dmitrienko}$^{1}$\sep
        } \runningauthor{I.S.
Dmitrienko} \runningtitle{FMS modes of the magnetotail }

   \institute{$^{1}$ Institute of Solar-Terrestrial Physics SB RAS, Lermontov St. 126,
Irkutsk 664033, Russia
                     email: \url{dmitrien@iszf.irk.ru} \\}
\begin{abstract}
FMS modes are studied in the model of the magnetotail as a
cylinder with  plasma sheet. The presence of the plasma sheet
leads to a significant modification of the modes existing in the
magnetotail in the form of a cylinder with no plasma sheet.
Azimuthal scales of the FMS modes differ significantly  between
the lobes and the plasma sheet. The azimuthal scale in the plasma
sheet  is much smaller than that in the magnetotail lobes. FMS
waves with certain parameters are strongly reflected from the
boundary between the lobes and the plasma sheet and are very weak
in the plasma sheet.

\end{abstract}

\end{opening}

\section{Introduction}
     \label{Section1}
The magnetotail is affected by various perturbations originating
 from the environment it is surrounded by. Such perturbations can
be transported by FMS waves across the magnetotail from areas of
their  origin.  FMS waves deep in the magnetotail are mode
converted  into Alfven waves, with the perturbation energy being
further transferred along the magnetic field by Alfven waves as
disturbances of various spatial scales \cite{0}. When FMS waves
are mode converted into Alfven waves, accelerated plasma flows are
formed \cite{1};  such flows can reach the surface of the planet.
The Alfven waves cause magnetic field perturbations and electron
precipitations due to their longitudinal electric field
\cite{3_1,3_2,3_3,3_4,2}. Thus,
 FMS waves of the magnetotail are not only of interest
as themselves but also in relation to the wide range of diverse
phenomena resulting from their mode conversion into Alfven waves.

Up until now the magnetotail FMS modes have been  studied in very
simple models only - either a cylindrical model without plasma
sheet \cite{4}, or a flat model \cite{5_1,5_2,5_3}. Despite their
simplicity, these models help capture some of the general features
in FMS wave propagation in the magnetotail. The first of the two
includes reflection of FMS waves propagating deep into the tail,
while the second model reflects the existence of an FMS waveguide
in the plasma sheet due to a strong inhomogeneity of the
magnetotail parameters in the direction across the magnetotail.
These results are, however, obtained in two different models, and
either model is clearly not satisfactory.  In this paper we
address FMS  modes in a model magnetotail taking into account both
the nearly cylindrical shape of the magnetotail  and the plasma
sheet within. The current sheet, with strongly attenuating
magnetic field inside, manifests itself as a strong inhomogeneity
of the Alfven velocity, in the FMS wave equations. Since the
dispersion properties of FMS waves in the area of existence of the
current sheet are determined by the sound velocity, whereas the
value of the low Alfven speed is insignificant, we neglect the
presence of the current sheet and confine ourselves to including
only the plasma sheet in the model.

We introduce a special orthogonal curvilinear coordinate system
which generalizes the radial and azimuthal cylindrical coordinates
for magnetotail cross-section. The boundary of the plasma sheet
and lobes  lies on the coordinate curves in this coordinate
system. In this case it is possible to reduce the problem of
describing the FMS modes to an ordinary second-order differential
equation by  using the  "radial " WKB approximation. Together with
the boundary conditions on the "azimuthal" coordinate, it gives a
dispersion equation that determines the wave "radial " numbers as
functions of "radial" coordinate.

\section{Equations for FMS waves in cilindrical madnetotail with plasma sheet.}
We assume that the magnetotail is a plasma column extended  along
the direction of magnetic field ${\bf B_{0}}$; this direction is
chosen as the $x$ coordinate of the Cartesian coordinate system.
All unperturbed plasma parameters and the magnetic field are
homogeneous with respect to this coordinate.

The homogeneity of the unperturbed parameters on $x$ lets us
consider a Fourier harmonic of the wave perturbation corresponding
to this coordinate, i.e. suppose  that the perturbation is the
product of functions $y$   and $z$ by $\exp(ikz)$, where $k$ is
the longitudinal wave number.  We will designate special
orthogonal coordinates in the  plane perpendicular to ${\bf
B_{0}}$ as $x^{1}$, $x^{2}$.   We have $g_3=1$, $\partial
_{3}g_{1}=\partial _{3}g_{2}=0$, $B_{03}=B_{0}^{3}=B_{0}$.

We will take linearized MHD equations:
\begin{equation}\label{eq:1}
\begin{array}{l}
\partial_{t}{\bf B}=\nabla \times\left[ {\bf v}\times{\bf
B}_{0}\right],\\ m_{i}n_{0}\partial _{t}{\bf v}
=\frac{1}{4\pi }\left[ \nabla \times {\bf B}\right]\times {\bf B}_{0} +\frac{1}{4\pi }\left[ \nabla \times{\bf B}_{0}\right]{\bf B} -{\bf \nabla }P,\\
\partial _{t}P+\left( {\bf v}\cdot\nabla \right) P_{0}+V_{s}^{2}\rho _{0}\nabla\cdot{\bf v}=0, \\
\partial_{t}\rho +\left( {\bf v}\cdot\nabla \right)
\rho_{0}+\rho_{0}\nabla\cdot{\bf v}=0. \\
\end{array}
\end{equation}
Here ${\bf B},{\bf v},\rho ,P$   are  magnetic field, velocity,
density and pressure, respectively. Unperturbed parameters are
subscripted with $0$; $V_{s}$ is sound velocity,
$V_{s}=\sqrt{\frac{\gamma P_{0}}{\rho _{0}}}$, where  $\gamma$  is
the adiabatic index.

We will designate  the components of the field and velocity
vectors  in the plane perpendicular to the ${\bf B}_0$ as $
B_{\perp k}, v_{\perp k}$; $k=1,2$.

Unperturbed parameters must satisfy the equations \\
$\partial _{k}\left( \frac{B_{0}^{2}}{8\pi }+P_{0}\right) =0$, \\
where $P_{0}=P_{0}\left( x^{1}\ ,x^{2}\right)$. We will put
\begin{equation}\label{eq:2}
P_{0}\left( x^{1}\ ,x^{2}\right)
=\overline{P}-\frac{B_{0}^{2}\left( x^{1},x^{2}\right) }{8\pi },
\end{equation}
where $\overline{P}=const$.

 Let us denote  the perturbation of the full pressure  as $\Psi$:
$\Psi =\frac{1}{4\pi }B_{0}B_{3}+P$. Taking into account
(\ref{eq:2}) the equation for $\Psi$ can be obtain from
(\ref{eq:1}) in the form
\begin{equation}\label{eq:8}
\partial _{1}\left(
\sqrt{\frac{g_{2}}{g_{1}}}\frac{1}{B_{0}^{2}l_{a}}\partial
_{1}\Psi \right) +\partial _{2}\left(
\sqrt{\frac{g_{1}}{g_{2}}}\frac{1}{B_{0}^{2}l_{a}}\partial
_{2}\Psi \right) +\sqrt{g}\frac{1}{B_{0}^{2}l_{a}}U\Psi =0,
\end{equation}
where $U=\frac{\left( \omega ^{2}V_{a}^{-2}-k^{2}\right) \left(
\omega ^{2}c_{s}^{-2}-k^{2}\right) }{\omega
^{2}V_{c}^{-2}-k^{2}}$, and $v_c$ is the velocity of slow
magnetosonic wave, $V_{c}=\frac{c_{s}V_{a}}{\sqrt{\left(
V_{a}^{2}+c_{s}^{2}\right) }}$.

We will use the approximation of homogeneous plasma in a uniform
field for the magnetotail lobes (its parts outside the plasma
layer up to the outer boundary). We will index the unperturbed
parameters of the lobes by the $l$ subscript. The lobe plasma is
cold $\left( \beta <<1\right) $, therefore
$V_{al}^{2}>>V_{sl}^{2}$ and we have $U=\omega ^{2}V_{al\
}^{-2}-k^{2}$. We will index the unperturbed parameters of the
plasma sheet by the  $p$-subscript. We will use the hot plasma
approximation for the plasma sheet and confine ourselves to the
homogeneous plasma and  field approximation. When $\beta >>1$,
therefore $V_{ap}^{2}<<V_{sp}^{2}$ and we have $U=\omega
^{2}V_{sp}^{-2}-k^{2}$. Thus strong inhomogeneity of the field in
the current sheet is not important for $U$. We obtain
 \begin{equation}\label{eq:9}
U=\begin{array}{l} \omega ^{2}V_{al}^{-2}-k^{2}, {\rm \ \ in \ the
\ lobes} \\
\omega ^{2}V_{sp}^{-2}-k^{2}, {\rm \ \ in \ the \ plasma {\rm \ }
  sheet}

\end{array}.
\end{equation}

We have $\frac{B_{0l}^{2}}{8\pi }+P_{0l}=\frac{B_{0p}^{2}}{8\pi
}+P_{0p}$ from (\ref{eq:2}). This equality can be replaced by the
approximate  equality $\frac{B_{0l}^{2}}{8\pi }=P_{0p}$ given the
inequalities $B_{0l}^{2}>>B_{0ps}^{2}$ and $P_{0l}<<P_{0p}$. Thus
we have $V_{sp}^{2}=V_{al\ }^{2}\frac{\gamma \rho _{0l}}{2\rho
_{0ps}}$ in (\ref{eq:9}).

Now let us specify the magnetotail cross section structure. We
assume it to consist of a plasma layer shaped as a rectangle with
sides $2h$ and $2R$ $(R>h)$ and two lobes shaped as semicircles
with  radius $R$ , their diameters adjacent to the longer side of
the plasma layer. We define  the coordinates in the plane as:
\begin{equation}\label{eq:10}
 \rho=
\begin{array}{c}
  \left\vert y\right\vert , {\rm \ when \ } \left\vert z\right\vert \leqslant h \\
  \sqrt{y^{2}+\left( \left\vert z\right\vert -h\right) ^{2}},  {\rm \ when \ } \left\vert z\right\vert >h\\
\end{array},
\end{equation}
\begin{equation}\label{eq:11}
 s=
\begin{array}{c}
  z, {\rm \ when \ } -h\leqslant z\leqslant h,{\rm \ }y>0 \\
  h+R\arccos \frac{y}{\sqrt{y^{2}+\left( \left\vert z\right\vert -h\right) ^{2}}}, {\rm \ when \ } z>h
  \\
  -h-R\arccos \frac{y}{\sqrt{y^{2}+\left( \left\vert z\right\vert -h\right) ^{2}}} , {\rm \ when \ } z<h\\
  S-z,{\rm \ when \ } 0<z\leqslant h,{\rm \ }y<0 \\
-S-z,{\rm \ when \ } 0>z\geqslant h,{\rm \ }y<0 \\
\end{array}.
\end{equation}

 Metric coefficients in the coordinate system : $x^{1}=\rho $, $x^{2}=s$ according to (\ref{eq:10}) and
 (\ref{eq:11}):\\
$
 g_1=1, g_2=\begin{array}{l}
   1, {\rm \ for \ } -h\leqslant z\leqslant h {\rm \ (the \ plasma \ sheet)}\\
   \left( \rho /R\right) ^{2}, {\rm \ for \ } z>h {\rm \ and \ } z<-h {\rm \ (the \ lobes)} \\
 \end{array}$. \\
In the $\rho,s$ coordinates, (\ref{eq:8}) takes the form
\begin{equation}\label{eq:13}
\frac{1}{\sqrt{g_{2}}}\partial _{\rho }\left( \sqrt{g_{2}}\partial
_{\rho }\Psi \right) +\frac{1}{\sqrt{g_{2}}}B_{0}^{2}l_{a}\partial
_{s}\left( \frac{1}{\sqrt{g_{2}}}\frac{1}{B_{0}^{2}l_{a}}\partial
_{s}\Psi \right) +U\Psi =0,
\end{equation}
 The parameters $B_0$, $\rho_0$ have jumps as functions of $s$ at
the boundary of the plasma sheet and the lobes; $g_2$ has a jump
as well.

The  integration of (\ref{eq:13}) over $s$  near the plasma
sheet-lobes boundary gives conditions for matching the solution at
this boundary as
\begin{equation}\label{eq:14}
\Psi \left( \overline{s}+0\right) =\Psi \left(
\overline{s}-0\right) ,
\end{equation}
\begin{equation}\label{eq:15}
\left[ \frac{1}{\sqrt{g_{2}}B_{0}^{2}l_{a}}\partial _{s}\Psi
\right] _{\left( \overline{s}+0\right) }-\left[
\frac{1}{\sqrt{g_{2}}B_{0}^{2}l_{a}}\partial _{s}\Psi \right]
_{\left( \overline{s}-0\right) }=0,
\end{equation}
where  $\overline{s}$  is  one of the two values of  $s$  at the
boundary given by the equations  $s=\pm h$.

Outside the plasma sheet-lobe boundary, equation (\ref{eq:13})
takes the form
\begin{equation}\label{eq:17}
\frac{1}{\sqrt{g_{2}}}\partial _{\rho }\left( \sqrt{g_{2}}\partial
_{\rho }\Psi \right) +\frac{1}{g_{2}}\partial _{s}\partial
_{s}\Psi +U\Psi =0.
\end{equation}
We will use the WKB approximation to solve
(\ref{eq:17}):\\
$\Psi =\psi _{+}\left( \rho ,s\right) \exp \left( i\int Kd\rho
\right) +\psi _{-}\left( \rho ,s\right) \exp \left( -i\int Kd\rho
\right) $,\\
where   $K$  is a  large wave number ($|K|<<R^{-1}$), which gives
a rapid change of $\Psi$ in the $\rho$   coordinate, while
  $\psi_+$  and $\psi_-$     are slowly varying functions of $\rho :\left\vert \partial _{\rho }\psi_{\pm} /\psi_{\pm} \right\vert <<\left\vert K\right\vert
  $.  The $+$  and $-$ subscripts of  $\psi$  will be omitted.

  We obtain from (\ref{eq:17}) in the main order of the WKB approximation:
\begin{equation}\label{eq:18}
\frac{1}{g_{2}}\partial _{s}\partial _{s}\psi +\left(
U-K^{2}\right) \psi =0.
\end{equation}
 Boundary conditions in the  $s$ coordinate are the periodicity conditions:
\begin{equation}\label{eq:19}
\psi (-S)=\psi (S), {\rm \ \ }\partial _{s}\psi (-S)=\partial
_{s}\psi (S).
\end{equation}

\section{Dispersion equation for FMS modes.}

 Taking into consideratuin the symmetry with respect to $s=0$  the periodicity conditions (\ref{eq:19}) on $[-S,+S]$ should be
replaced by the following two boundary conditions within $[-S,0]$:
\begin{equation}\label{eq:20}
\partial _{s}\psi \left( 0\right) =0, {\rm \ \ }\partial _{s}\psi \left(
-S\right) =0
\end{equation}
and
\begin{equation}\label{eq:21}
\psi \left( 0\right) =0,{\rm \ \ }\psi \left( -S\right) =0.
\end{equation}

Since there is symmetry  on the  $[-S,0]$ with respect to $s=-S/2$
we can proceed further to the boundary conditions on $[-S,-S/2]$
and obtain:
\begin{equation}\label{eq:29}
\partial _{s}\psi (-S)=0,{\rm \ \ }\partial _{s}\psi (-S/2)=0,
\end{equation}
\begin{equation}\label{eq:30}
\psi (-S)=0,{\rm \ \ }\partial _{s}\psi (-S/2)=0,
\end{equation}
\begin{equation}\label{eq:31}
\partial _{s}\psi (-S)=0,{\rm \ \ }\psi (-S/2)=0,
\end{equation}
\begin{equation}\label{eq:32}
\psi (-S)=0,{\rm \ \ }\psi (-S/2)=0,
\end{equation}
instead of (\ref{eq:20}) and (\ref{eq:21}).

In the case of boundary conditions (\ref{eq:29}), we will seek the
solution to (\ref{eq:18}) in the plasma sheet in the form
\begin{equation}\label{eq:33}
\psi =A\cos \left[ Q_{p}\left( S+s\right) \right] ,{\rm \ \
}Q_{p}=\sqrt{ U_{p}-K^{2} \ },{\rm \ \ }U_{p}=\ \omega
^{2}V_{sp}^{-2}-k^{2};
\end{equation}
We will seek the solution in the lobes as
\begin{equation}\label{eq:34}
\psi =a\cos \left[ Q_{l\ }\left( S/2+s\right) \right] ,{\rm \ \
}Q_{l\ }=\frac{\rho }{R}\sqrt{ U_{l\ }-K^{2} \ },{\rm \ \ }U_{l\
}=\omega ^{2}V_{al}^{-2}-k^{2}.
\end{equation}
These solutions satisfy (\ref{eq:29}) and we only have to match
them at $s=-S+h$. To obtain the dispersion equation it is
convenient to use the matching of the logarithmic derivative
instead of (\ref{eq:14}), (\ref{eq:15}):
\begin{equation}\label{eq:35}
\left[ \frac{1}{\sqrt{g_{2}}B_{0}^{2}l_{a}}\frac{\partial _{s}\Psi
}{\Psi }\right] _{\left( \overline{s}+0\right) }-\left[
\frac{1}{\sqrt{g_{2}}B_{0}^{2}l_{a}}\frac{\partial _{s}\Psi }{\Psi
}\right] _{\left( \overline{s}-0\right) }=0.
\end{equation}
Substituting (\ref{eq:33}) and (\ref{eq:34}) in (\ref{eq:35}) we
get
\begin{equation}\label{eq:36}
\tan \left[ Q_{l\ }\left( -S/2+h\right) \right] =\delta ^{-1}\tan
\left[ Q_{p}h\right] .
\end{equation}
We denoted $\delta =\frac{\sqrt{ U_{l\ }-K^{2} \ }\left(
B_{0}^{2}l_{a}\right) _{p}}{\sqrt{ U_{p}-K^{2} \ }\left(
B_{0}^{2}l_{a}\right) _{l}}$. The ratio of the mode amplitudes in
the plasma sheet to those in the lobes is determined from
(\ref{eq:15}):
\begin{equation}\label{eq:37}
\frac{A}{a}=\delta \frac{\sin \left[ Q_{l\ }\left( -S/2+h\right)
\right] }{\sin \left[ Q_{p}h\right] }.
\end{equation}
The dispersion equation and the amplitude ratio for  (\ref{eq:30})
- (\ref{eq:32}) are obtained similarly:
\begin{equation}\label{eq:38}
\tan \left[ Q_{l\ }\left( -S/2+h\right) \right] =-\delta ^{-1}\tan
^{-1}\left[ Q_{p}h\right] , {\rm \ \ }\ \frac{A}{a}=-\delta
\frac{\sin \left[ Q_{l\ }\left( -S/2+h\right) \right] }{\cos
\left[ Q_{ps}h\right] }
\end{equation}
for  (\ref{eq:30}),
\begin{equation}\label{eq:39}
\tan \left[ Q_{l\ }\left( -S/2+h\right) \right] =-\delta \tan
^{-1}\left[ Q_{p}h\right] , {\rm \ \ }\frac{A}{a}=-\delta
\frac{\cos \left[ Q_{l\ }\left( -S/2+h\right) \right] }{\sin
\left[ Q_{p}h\right] }
\end{equation}
for  (\ref{eq:31}),
\begin{equation}\label{eq:40}
\tan \left[ Q_{l\ }\left( -S/2+h\right) \right] =\delta \tan
\left[ Q_{p}h\right] ,{\rm \ \ }\frac{A}{a}=\delta \frac{\cos
\left[ Q_{l\ }\left( -S/2+h\right) \right] }{\cos \left[
Q_{p}h\right] }
\end{equation}
for  (\ref{eq:32}).

  It is easy to see that when  $\delta \rightarrow \infty $ we generally get  values of the order of $1$
  for $A/a$ from (\ref{eq:37}) - (\ref{eq:40}). This means that the amplitudes of
  the modes in the plasma sheet and in the lobes are values of the same order for large values of  $\delta$.

  When   $\delta \rightarrow 0$   it follows from  (\ref{eq:37})-(\ref{eq:40}) that  $\frac{A}{a}\sim \delta
  $.
       A small ratio  $\frac{A}{a}$ means that for small  $\delta$ the modes
  are almost not present in the plasma sheet and are concentrated in the magnetotail lobes.  This means that
  the FMS modes propagating in the lobes are strongly  reflected  from their boundary with the plasma sheet.

The results of numerical calculations  of  the eigenvalues
$p^2(r)$ and the corresponding ratio of the amplitudes as well as
$q_p(r)$, $q_l(r)$ and $\delta(r)$ for (\ref{eq:36}) are shown in
Fig. \ref{fig:3}. Fig. \ref{fig:3} is the case of parameters for
which the modes decay from the magnetotail boundary deeper into
the magnetotail. Fig. \ref{fig:3}a shows  $p^2(r)$, while Fig.
\ref{fig:3}b and \ref{fig:3}c show wave numbers $q_p(r)$ and
$q_l(r)$ corresponding to the coordinate $s$. We see in figures
\ref{fig:3}b and \ref{fig:3}c that azimuthal scales of the modes
are substantially smaller in the plasma sheet than those in the
lobes. Fig. \ref{fig:3}d depicts $\delta(r)$. As we can see,
$\delta$  is small. The curves in Fig. \ref{fig:3}d  show that the
modes are mainly concentrated in the magnetotail lobes.
\begin{figure}\label{fig:3}
\includegraphics[width=1\linewidth]{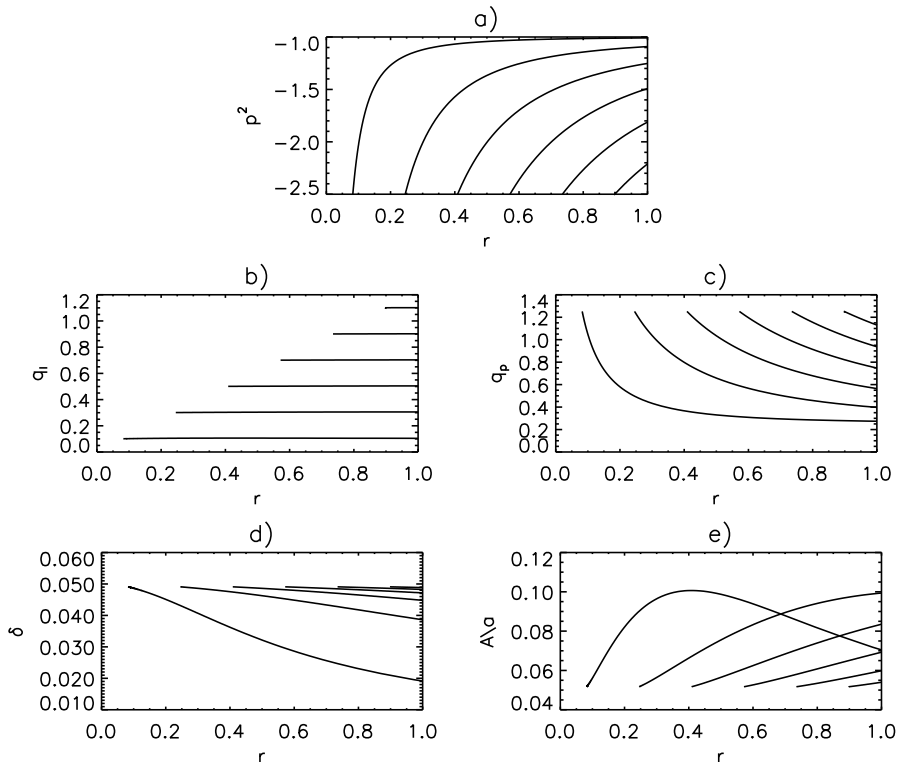}\\
\caption{Plots of  numerically calculated  $p^{2}\left( r\right)$,
$q_{l}$ $\left( r\right)$, $q_{p}\left( r\right)$, $\delta \left(
r\right)$,
$\frac{A}{a}\left( r\right) $ when $h/R=0.1$,$b=0.1$, $\lambda =0.05$, $w=0.05$, $t=10$, $v=0.05$. %
 }\label{fig:3}
\end{figure}

\section{Conclusions}
A model of the magnetotail  is proposed taking into account both
its external structure and the presence of a plasma sheet within.
A special system of coordinates is introduced corresponding to
such a model. Dispersion equations for FMS modes are derived using
the model and the coordinates introduced. It is shown that the
azimuthal scales of FMS modes are substantially different between
the lobes and the plasma sheet - the azimuthal scale in the plasma
sheet  is much smaller than that in the magnetotail lobes; the
mode amplitude distributions in the azimuthal coordinate are
inhomogeneous.  The FMS waves decaying deep into the magnetotail
but propagating along the azimuth coordinate are primarily
concentrated in the lobes and weakly penetrate into the plasma
sheet.


\end{article}

\end{document}